\documentclass[12pt]{article}
\pdfoutput=1

\usepackage{amsmath,amssymb,amscd}
\usepackage{listings}
\usepackage{caption}
\usepackage{dsfont}
\usepackage{slashed}
\usepackage{color}
\usepackage{comment}
\usepackage[pdftex]{graphicx}
\usepackage{epstopdf}
\usepackage{subfigure}
\usepackage{epsfig}
\usepackage{listings}
\usepackage{caption}
\usepackage{cite}
\usepackage{hyperref}

\usepackage{multirow}
\usepackage[titletoc]{appendix}

\usepackage{array}
\newcolumntype{M}[1]{>{\centering\arraybackslash}m{#1}}
\newcolumntype{N}{@{}m{0pt}@{}}

\allowdisplaybreaks

\newcommand{\half}{\frac{1}{2}}
\newcommand{\bea}{\begin{align}}
\newcommand{\eea}{\end{align}}
\newcommand{\beq}{\begin{equation}}
\newcommand{\eeq}{\end{equation}}
\newcommand{\nbea}{\begin{align*}}
\newcommand{\neea}{\end{align*}}
\newcommand{\nbeq}{\begin{equation*}}
\newcommand{\neeq}{\end{equation*}}
\newcommand{\bear}{\begin{eqnarray}}  
\newcommand{\eear}{\end{eqnarray}}  

 \newcommand{\twomatrix}[1]{\left(\begin{array}{cc} #1 \end{array}\right) }
  \newcommand{\threematrix}[1]{\left(\begin{array}{ccc} #1 \end{array}\right) }
 \newcommand{\column}[1]{\left(\begin{array}{c} #1 \end{array}\right) }
  \newcommand{\inviscolumn}[1]{\begin{array}{c} #1 \end{array} }
 
 \newcommand{\identity}{\mathds{1}}

\newcommand{\Dfbu}{\mathord{\buildrel{\lower3pt\hbox{$\scriptscriptstyle{\leftrightarrow \tiny{ \ \ \ } }$}}\over {D^{\mu}}}} 
\newcommand{\Dfbd}{\mathord{\buildrel{\lower3pt\hbox{$\scriptscriptstyle\longleftrightarrow$}}\over {D}_{\mu}}} 

\numberwithin{equation}{section}

\newcommand{\Htil}{\tilde{H}}
\newcommand{\Hdtil}{\tilde{H}^\dagger}
\newcommand{\Hd}{H^\dagger}
\newcommand{\Dm}{D_\mu}
\newcommand{\Dv}{\text{div.}}

\begin{document}

\begin{titlepage}

\pagestyle{empty}

\baselineskip=21pt
\rightline{\small MCTP-16-08, KCL-PH-TH/2016-18, Cavendish-HEP-16/04, DAMTP-2016-31}
\vskip 0.6in

\begin{center}

{\large {\bf  Mixed Heavy-Light Matching in the Universal One-Loop Effective Action}}

\vskip 0.4in

 {\bf Sebastian~A.~R.~Ellis}$^{1}$,
 {\bf J\'er\'emie~Quevillon}$^{2}$,
{\bf Tevong~You}$^{3}$
and {\bf Zhengkang~Zhang}$^{1}$

\vskip 0.4in

{\small {\it
$^1${Michigan Center for Theoretical Physics (MCTP), \\ Department of Physics, University of Michigan \\Ann Arbor, MI 48109 USA}\\
\vspace{0.25cm}
$^2${Theoretical Particle Physics and Cosmology Group, Physics Department, \\
King's College London, London WC2R 2LS, UK}\\
\vspace{0.25cm}
$^3${DAMTP, University of Cambridge, Wilberforce Road, Cambridge, CB3 0WA, UK; \\
Cavendish Laboratory, University of Cambridge, J.J. Thomson Avenue, \\ 
\vspace{-0.25cm}
Cambridge, CB3 0HE, UK}
}}

\vskip 0.5in

{\bf Abstract}

\end{center}

\baselineskip=18pt \noindent


{\small
Recently, a general result for evaluating the path integral at one loop was obtained in the form of the Universal One-Loop Effective Action. It may be used to derive effective field theory operators of dimensions up to six, by evaluating the traces of matrices in this expression, with the mass dependence encapsulated in the universal coefficients. Here we show that it can account for loops of mixed heavy-light particles in the matching procedure. Our prescription for computing these mixed contributions to the Wilson coefficients is conceptually simple. Moreover it has the advantage of maintaining the universal structure of the effective action, which we illustrate using the example of integrating out a heavy electroweak triplet scalar coupling to a light Higgs doublet. Finally we also identify new structures that were previously neglected in the universal results. 
}


\vskip 1in

{\small \leftline{April 2016}}

\end{titlepage}

\newpage



\section{Introduction}

Matching from an ultraviolet (UV) theory to a low-energy effective field theory (EFT) can be performed using either Feynman diagrams or functional methods. For the latter approach, Gaillard~\cite{Gaillard} and Cheyette~\cite{Cheyette} introduced a manifestly gauge-covariant method of performing the calculation, using a covariant derivative expansion (CDE). This elegant method simplifies evaluating the quadratic term of the heavy fields in the path integral to obtain the low-energy EFT, and was revived recently by Henning, Lu and Murayama (HLM)~\cite{HLM}. In particular, HLM pointed out that under the assumption of degenerate particle masses they could evaluate the momentum dependence of the coefficients that factored out of the trace over the operator matrix structure, without specifying the specific UV model. In Ref.~\cite{UOLEA} some of us showed that this universality property can be extended without any assumptions on the mass spectrum, to obtain a universal result for the one-loop effective action for operators up to dimension six. There the loop integrals have been computed for a general mass spectrum once and for all. This Universal One-Loop Effective Action (UOLEA) is a general expression that may then be applied in any context where a one-loop path integral needs to be computed, as for example in matching new physics models to the Standard Model (SM) EFT~\footnote{For recent matching calculations see for example Refs.~\cite{HLM, UOLEA, HKMN, HLMstops, GNS, DEQYstops, chianghuo, huo1, huo2, AKS}. The SM EFT is reviewed in Refs.~\cite{SMEFTreview1, SMEFTreview2}.}.

Functional methods require the term quadratic in the heavy fields to be integrated out, corresponding to loops of heavy fields with light particle external legs in the Feynman diagram approach. In addition to these heavy-heavy loops, there could also be mixed heavy-light contributions to matching. These are typically calculated using Feynman diagrams~\cite{bilenkysantamaria, skiba, skibatasi, AKS} but can also be accounted for in the functional approach~\cite{mixedfunctional1,mixedfunctional2, BGP}. The purpose of this paper is to show how they can be computed in the UOLEA. 

Compared to previous functional methods~\cite{mixedfunctional1,mixedfunctional2, BGP}, our prescription for treating mixed heavy-light contributions is relatively simple and transparent: in addition to the usual expansion of the heavy fields around their classical solution, we also separate the light fields into classical and quantum parts, and extend the quadratic term to also include quantum fluctuations of the light fields.
This essentially amounts to computing the 1PI effective action for the full theory, from which the Wilsonian effective Lagrangian, namely the low-energy EFT, can be extracted. Similarly to the heavy-heavy case, the general structure and universal coefficients of the UOLEA combine to yield the EFT Wilson coefficients after evaluating the matrix traces. But in this extended case, the universal coefficients contain parts that are in the full 1PI effective action but not in the EFT, diagrammatically corresponding to tree-generated operator insertions in EFT loops. These must be subtracted by a well-defined procedure, which we describe. 
Our prescription has the advantage of maintaining the universal structure of the UOLEA so that in principle, one need not apply the CDE  starting from the beginning for every model.

We also find that in certain cases, for example when including vector gauge boson contributions, the matrix structure may contain an extra covariant derivative part that is not taken into account in the pre-evaluated form of the UOLEA; Refs.~\cite{HLM, UOLEA} assume no such additional structure in its derivation. These new contributions then have to be computed separately for each specific case using the CDE method to evaluate the path integral from the beginning. However, it is possible in principle to do the calculation in a model-independent way, once and for all, which would extend the UOLEA to include such structures. Such an extension will be addressed in future work~\cite{workinprogress}.

In the next Section we give a brief introduction to the CDE method and the UOLEA. In Section~\ref{sec:mixedheavylight} we outline the procedure for including mixed heavy-light contributions to dimension-6 operators with the UOLEA. As an example, in Section~\ref{sec:electroweaktriplet} we demonstrate how to obtain heavy-light contributions to matching a heavy electroweak triplet scalar model to the SM EFT, and discuss the extension needed to incorporate gauge coupling-dependent contributions.
Finally we conclude in Section~\ref{sec:conclusion}. Some useful formulae are collected in the Appendix.

\section{The Universal One-Loop Effective Action}
\label{sec:UOLEA}

We begin by describing the Gaillard-Cheyette Covariant Derivative Expansion (CDE) method~\cite{Gaillard, Cheyette} for evaluating the path integral~\footnote{See Ref.~\cite{HLM} for a review and more technical details.}. The UV Lagrangian for a model composed of light and heavy fields, that we collectively denote as the multiplets $\phi$ and $\Phi$ respectively, can be written as 
\begin{equation}
\mathcal{L}_\text{UV}[\phi,\Phi] \supset \mathcal{L}[\phi] + \Phi\cdot F[\phi] + \half\Phi(P^2 - \mathcal{M}^2 - U^\prime[\phi])\Phi + \mathcal{O}(\Phi^3) \, ,
\label{eq:lagrangianUV}
\end{equation}
where $\mathcal{L}[\phi]$ is the light field part of the Lagrangian and the gauge-covariant derivative $D_\mu$ is written as $P_\mu \equiv iD_\mu$. $\mathcal{M}$ is a diagonal mass matrix. Eq.~\ref{eq:lagrangianUV} is written for a real scalar $\Phi$; in general the exact form depends on the nature of $\Phi$. The terms involving light fields coupling linearly and quadratically to $\Phi$ are represented by the matrices $F[\phi]$ and $U^\prime[\phi]$ respectively. 

Beginning from an action $S[\phi,\Phi]$, we can expand around the minimum and evaluate the path integral over $\Phi$. For example in the case of real scalar fields the effective action can be written as
\begin{align*}
e^{iS_\text{eff}[\phi]} &= \int [D\Phi] e^{iS[\phi,\Phi]} \\
&= \int [D\eta] e^{i\left(S[\phi,\Phi_c] + \frac{1}{2}\left.\frac{\delta^2 S}{\delta \Phi^2}\right|_{\Phi=\Phi_c}\eta^2 + \mathcal{O}(\eta^3)\right)} \\
&\approx e^{iS[\phi,\Phi_c]}\left[\text{det}\left(\left.-\frac{\delta^2 S}{\delta\Phi^2}\right|_{\Phi=\Phi_c}\right)\right]^{-\frac{1}{2}}	\\
&= e^{iS[\phi,\Phi_c] - \frac{1}{2}\text{Tr ln}\left(-\left.\frac{\delta^2 S}{\delta \Phi}\right|_{\Phi=\Phi_c}\right)} \, ,
\end{align*}
%
where we used $\Phi= \Phi_c + \eta$ and we have defined $\Phi_c$ as the classical solution to $\left.\frac{\delta S}{\delta \Phi}\right|_{\Phi = \Phi_c} = 0$. This is applicable to bosons or fermions. In general the result is a one-loop effective action of the form 
\begin{equation}
S_{\text{1-loop}}^{\text{eff}} = i c_s \text{Tr ln}\left( -P^2 + \mathcal{M}^2 + U \right)	\, . 
\label{eq:S1loop}
\end{equation}
The constant $c_{s}$ depends on the heavy field $\Phi$. If it is a real scalar, complex scalar, 
Dirac fermion, gauge boson or Fadeev-Popov ghost then it takes the value $1/2,1,-1/2,1/2$ or $-1$ respectively~\cite{HLM}. We note that the $U$ matrix in Eq.~\ref{eq:S1loop} is obtained after a suitable rearrangement to the required form. The relation of $U$ to the quadratic term $U^{\prime}$ of the original Lagrangian depends on the species of $\Phi$, i.e.\ on whether we are dealing with a real or complex scalar, fermion, gauge boson, and so on. For more details we refer the reader to Ref.~\cite{HLM}. As we will see later Refs.~\cite{HLM, UOLEA} have the implicit assumption that $U$ does not contain any covariant derivatives acting openly to the right. 

After evaluating the trace over spacetime by inserting a complete set of spatial and momentum eigenstates, we have a trace ``tr'' over internal indices (gauge, flavour, spinor, etc.):
\begin{equation*}
S_{\text{1-loop}}^{\text{eff}} = i c_s\int d^dx \int\frac{d^dq}{(2\pi)^d} \text{tr ln}\left( -(P_\mu - q_\mu)^2 + \mathcal{M}^2 + U \right)	\, , 
\end{equation*}
where $d=4-\epsilon$ in dimensional regularization. Before manipulating the logarithm to obtain an expansion in terms of higher dimension operators, we shift the momentum in the integral using the covariant derivative by inserting factors of $e^{\pm P_\mu\partial/\partial q_\mu}$: 
\begin{equation*}
\mathcal{L}^\text{eff}_\text{1-loop} = i c_{s} \int \frac{d^dq}{(2\pi)^d} \text{tr} \ln[e^{P_\mu\partial/\partial q_\mu}(-(P_\mu-q_\mu)^2 + \mathcal{M}^2 + U) e^{-P_\mu\partial/\partial q_\mu} ] \, .
\end{equation*}
This ensures that $P_\mu$'s only appear in commutators, and the expansion will only involve manifestly 
gauge-covariant pieces throughout | that is the gauge field strengths, 
covariant derivatives and the SM fields encoded in the matrix $U(x)$:
\begin{equation}
\mathcal{L}^\text{eff}_\text{1-loop} = i c_{s} \int \frac{d^dq}{(2\pi)^d} \text{tr} \ln[-(\tilde{G}_{\nu\mu}\frac{\partial}{\partial q_\nu} + q_\mu)^2 + \mathcal{M}^2 + \tilde{U}]	\, ,
\label{Leff_ln}
\end{equation}
where
\begin{align*}
\tilde{G}_{\nu\mu} &\equiv \sum_{n=0}^{\infty} \frac{n+1}{(n+2)!}[P_{\alpha_1},[...[P_{\alpha_n},G^\prime_{\nu\mu}]]]\frac{\partial^n}{\partial q_{\alpha_1} ... q_{\alpha_n}} 	\, , \\
\tilde{U} &\equiv\sum_{n=0}^{\infty} \frac{1}{n!}[P_{\alpha_1},[...[P_{\alpha_n},U]]]	\frac{\partial^n}{\partial q_{\alpha_1} ... q_{\alpha_n}} \, ,
\end{align*}
and we defined $G^\prime_{\nu\mu}$ as the field strength given by 
$[P_\nu,P_\mu] = -G^\prime_{\nu\mu}$. 
This covariant formulation is the essence of the CDE method. 

In order to obtain the coefficients and structure of the higher dimension operators, there are various approaches one can take. For degenerate masses one can easily expand the action in Eq. (\ref{Leff_ln}) by integrating once its derivative with respect to the common mass scale $m^2$, as discussed in \cite{HLM}, or by making use of the Baker-Campbell-Hausdorff (BCH) formula as in \cite{Cheyette, DEQYstops}. However, for the general case of possibly non-degenerate masses, the mass matrix no longer commutes with the other matrix structures and the factorisation of the momentum integral from this structure is no longer trivial. To perform the expansion, one may use the BCH, or introduce an auxiliary parameter $\xi$ that multiplies the diagonal mass matrix $\mathcal{M}$, defined as
\beq
\mathcal{M} = \xi \cdot \text{Diag}(m_i) \ ,
\eeq
that can now be differentiated with respect to and integrated over. After the integration, we set $\xi = 1$. In the non-degenerate case, Eq. (\ref{Leff_ln}) is replaced by 
\beq
\mathcal{L}^\text{eff}_\text{1-loop} = -i c_{s} \int \frac{d^dq}{(2\pi)^d} \int d \xi \ \text{tr} \left[ \frac{1}{\Delta_\xi^{-1} + \{ q^\mu,\ \tilde{G}_{\nu\mu}\} \frac{\partial}{\partial q_\nu} + \tilde{G}_{\sigma \mu} {\tilde{G}^{\sigma}}_{\,\,\,\nu}  \frac{\partial}{\partial q_\mu} \frac{\partial}{\partial q_\nu} - \tilde{U} }\mathcal{M}^2\right] \ , 
\eeq
and then Taylor expanded to give 
\beq
\mathcal{L}^\text{eff}_\text{1-loop} = -i c_{s} \int \frac{d^dq}{(2\pi)^d} \int d \xi \ \text{tr} \left\{ \sum_{n=0}^{\infty} \left[ -\Delta_\xi \left(\{ q^\mu,\ \tilde{G}_{\nu\mu}\}\frac{\partial}{\partial q_\nu} + \tilde{G}_{\sigma \mu} {\tilde{G}^{\sigma}}_{\,\,\,\nu}\frac{\partial}{\partial q_\mu} \frac{\partial}{\partial q_\nu} - \tilde{U}\right)\right]^n \Delta_\xi \mathcal{M}^2\right\}\ .
\eeq
The matrices $\mathcal{M}^2$, $\Delta_\xi \equiv 1/(q^2 - \xi \mathcal{M}^2)$ and $\tilde{G}$ will not necessarily commute with $\tilde{U}$.

 The UOLEA is the result of evaluating the integrals in this expansion, extracting the coefficients and operator structure, eventually giving the following expressions relevant for operators of dimensions up to six~\cite{UOLEA}: 
\begin{align}
{\cal L}^{\text{eff}}_{\text{1-loop}}[\phi] \supset -i c_s & \Bigg\{ 
f_{1}^{i} + f_{2}^{i}U_{ii} + f_{3}^{i} G_{\mu\nu, ij}'^2 + f_{4}^{ij} U_{ij}^2  \nonumber \\
& + f_{5}^{ij}(P_{\mu} G_{\mu \nu,ij}^{'})^{2} + f_{6}^{ij}(G_{\mu \nu,ij}^{'})(G_{\nu \sigma,jk}^{'}) (G_{\sigma \mu,ki}^{'}) +f_{7}^{ij} [P_{\mu}, U_{ij}]^2 
\nonumber \\
& 
+f_{8}^{ijk} (U_{ij}U_{jk}U_{ki})   + f_{9}^{ij} (U_{ij} G_{\mu\nu,jk}'G_{\mu\nu,ki}^{'}) \nonumber \\
&+ f_{10}^{ijkl} (U_{ij}U_{jk}U_{kl} U_{li}) + f_{11}^{ijk} U_{ij} [P_{\mu},U_{jk}][P_{\mu},U_{ki}]  \nonumber \\
& +f_{12,a}^{ij} \left[P_{\mu},[P_{\nu},U_{ij}]\right]\left[P_{\mu}, [P_{\nu},U_{ji}]\right] + f_{12,b}^{ij} \left[P_{\mu},[P_{\nu},U_{ij}]\right] \left[ P_{\nu}, [P_{\mu},U_{ji}]\right]  \nonumber \\
& + f_{12,c}^{ij} \left[P_{\mu},[P_{\mu},U_{ij}]\right]\left[P_{\nu},[P_{\nu},U_{ji}]\right]
\nonumber \\
& +f_{13}^{ijk} U_{ij} U_{jk} G_{\mu \nu,kl}^{'}G_{\mu \nu,li}^{'} + f_{14}^{ijk}\left[ P_{\mu},U_{ij}\right] \left[ P_{\nu},U_{jk}\right] G_{\nu \mu,ki}^{'}  \nonumber \\
& +\left(f_{15a}^{ijk} U_{i,j} [ P_{\mu},U_{j,k}] - f_{15b}^{ijk} [P_{\mu},U_{i,j}] U_{j,k} \right) [P_{\nu},G^{'}_{\nu\mu , ki}] \nonumber \\
& + f_{16}^{ijklm}(U_{ij}U_{jk}U_{kl} U_{lm} U_{mi}) + f_{17}^{ijkl} U_{ij} U_{jk}  [P_{\mu},U_{kl}][P_{\mu},U_{li}] 
\nonumber \\
& +f_{18}^{ijkl}  U_{ij} [P_{\mu},U_{jk}] U_{kl} [P_{\mu},U_{li}]  +f_{19}^{ijklmn} (U_{ij}U_{jk}U_{kl} U_{lm} U_{mn} U_{ni}) 
 \Bigg\} \, .
\label{eq:universallagrangian}
\end{align}
The indices $i,j,k,l,m,n$ range over the dimension of the mass matrix $\mathcal{M}$ using an implied summation convention for repeated indices, and the $f_{N}$ are the universal coefficients that encapsulate the mass parameter dependence from loop integrals over momenta. Explicit expressions for these can be found in Ref.~\cite{UOLEA}. In the degenerate mass limit Eq.~\ref{eq:universallagrangian} reduces to the result of Ref.~\cite{HLM}.

\section{Integrating out mixed heavy-light contributions}
\label{sec:mixedheavylight}

\begin{figure}[h!]
\centering
\includegraphics[scale=0.5]{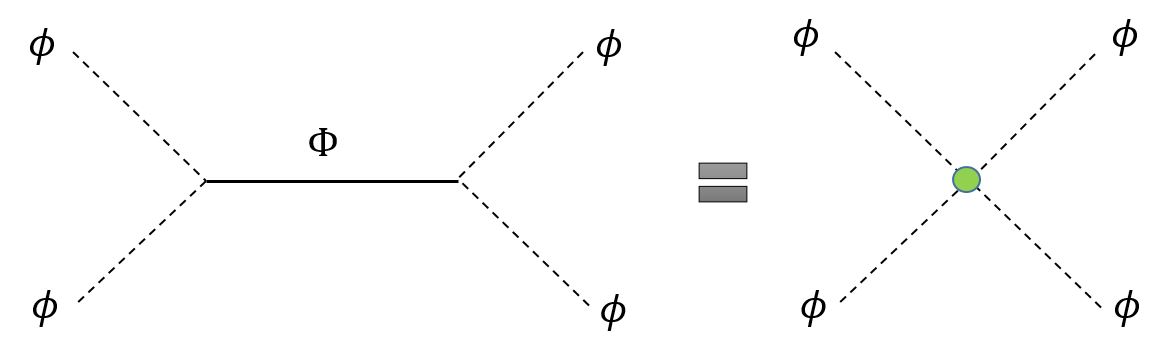}
\caption{\it Diagrammatic interpretation of tree-level matching with a light field $\phi$ coupling linearly to a heavy field $\Phi$ in the full UV theory on the left matched to the EFT local operator on the right. }
\label{fig:diagramtree}
\end{figure}

In the presence of light fields coupling linearly to the heavy particles, it initially appears that matching using the functional method does not account for mixed heavy-light contributions at one-loop, as argued for example in Refs.~\cite{bilenkysantamaria, AKS}. This is because the linear coupling in the UV Lagrangian of Eq.~\ref{eq:lagrangianUV} is responsible for the classical equation of motion of the heavy field given by 
\begin{equation}
\Phi_c \simeq \frac{F[\phi]}{-P^2 + \mathcal{M}^2 + U[\phi]} \simeq \frac{1}{\mathcal{M}^2}\sum^N_{n=0} \left( \frac{P^2-U[\phi]}{\mathcal{M}^2}\right)^n F[\phi] \, .
\end{equation}
The asymptotic expansion for the non-local operator $(-P^2 + \mathcal{M}^2 + U)^{-1}$ must be truncated at some finite order $N$, ensuring a local operator for the classical solution to be substituted back into the Lagrangian to integrate out the heavy particle at tree level, as depicted in Fig.~\ref{fig:diagramtree}. However, this procedure only suppresses the linear term by $\mathcal{M}^{-2N}$, and for any choice of $N$ they still contribute to order $\mathcal{M}^{-2}$ at one loop~\cite{AKS}. These correspond to the mixed heavy-light contributions neglected by the procedure for evaluating the quadratic term in the path integral outlined in the previous section. 

\begin{figure}[h!]
\centering
\includegraphics[scale=0.5]{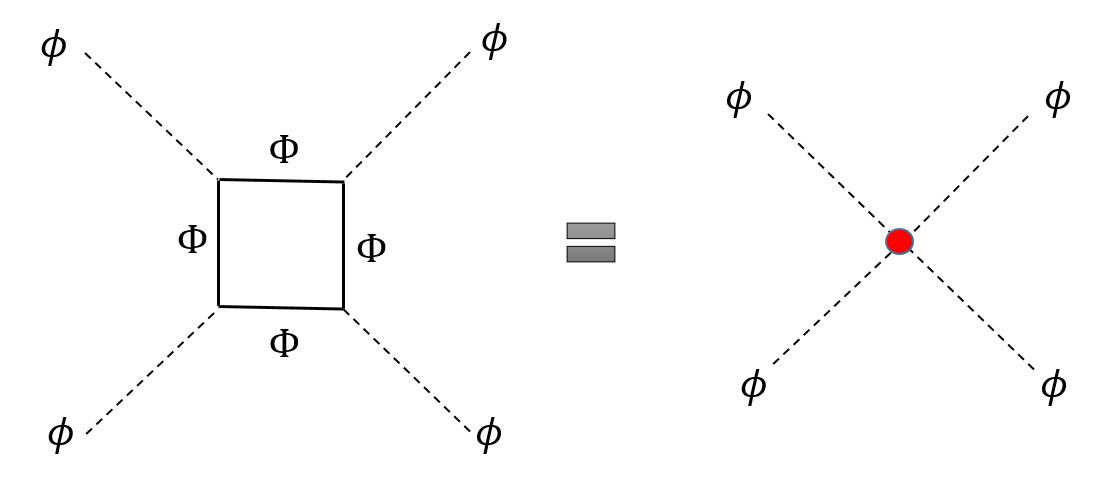}
\caption{\it Diagrammatic intepretation of one-loop matching with a light field $\phi$ coupling quadratically to a heavy field $\Phi$ in the full UV theory on the left matched to the EFT local operator on the right.  }
\label{fig:diagramoneloopheavyonly}
\end{figure}

\begin{figure}[h!]
\centering
\includegraphics[scale=0.5]{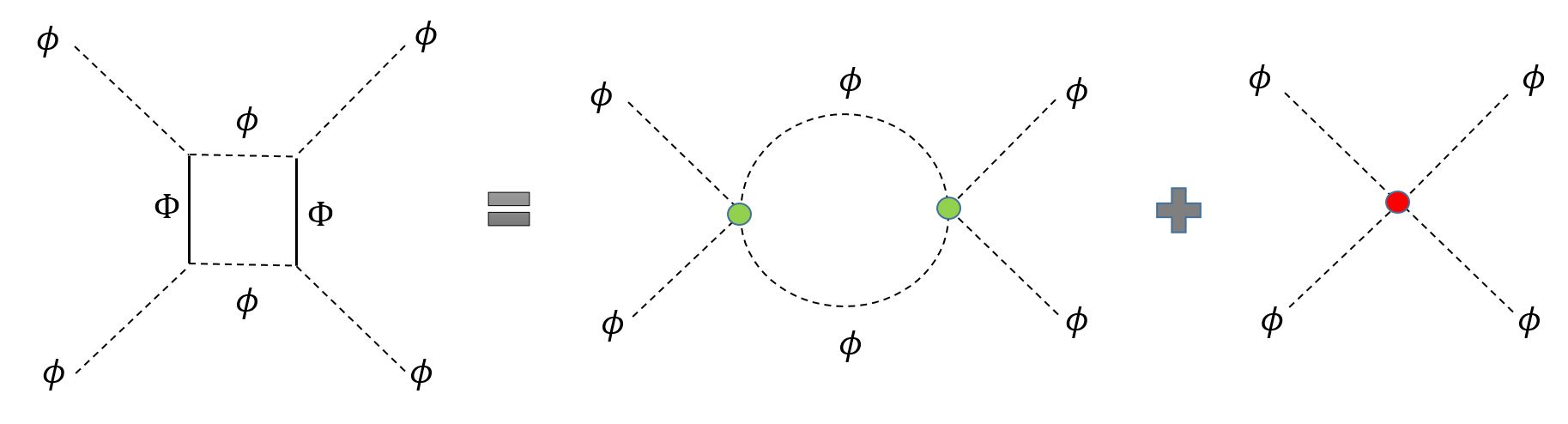}
\caption{\it Diagrammatic interpretation of one-loop matching with light and heavy fields $\phi$ and $\Phi$ respectively in the loop of the UV theory on the left, with the EFT contribution from tree-generated operator insertions used at one loop level and one-loop-generated local operators used at tree level on the right. }
\label{fig:diagramoneloop}
\end{figure}

An alternative, perhaps more intuitive way to understand this is based on the diagrammatic interpretation of the functional trace Eq.~\ref{eq:S1loop}. With the light fields treated as classical backgrounds, and only the heavy fields allowed to fluctuate, evaluating Eq.~\ref{eq:S1loop} essentially reproduces the sum of one-loop diagrams with heavy fields in the loop as illustrated in Fig.~\ref{fig:diagramoneloopheavyonly}. However, in the diagrammatic approach, local effective operators in the low-energy EFT also receive contributions from one-loop diagrams with both heavy and light fields in the loop. This is depicted in Fig.~\ref{fig:diagramoneloop}, where the UV diagram on the left is reproduced by the two EFT contributions on the right. It is the second contribution on the right that has not been captured in previous formulations of the functional approach to matching.

To include these mixed heavy-light loops, we follow the above diagrammatic intuition. In fact, we simply need to expand also the light fields $\phi$ with quantum fluctuations $\phi^\prime$ around their classical values $\phi_c$, in the same way as we did for the heavy fields $\Phi$,
\begin{equation}
\phi \to \phi_c + \phi^\prime \quad , \quad \Phi \to \Phi_c + \Phi^\prime \, ,
 \end{equation} 
where, to make the notation more transparent, we have replaced $\eta$ in Section~\ref{sec:UOLEA} by $\Phi'$. Substituting this into the UV Lagrangian we have an extended quadratic term for a multiplet involving the {\it quantum fluctuation parts} of the heavy and light fields together, of the form 
\begin{equation}
\mathcal{L}_\text{quad}  = \half\left({\Phi^\prime}, {\phi^\prime}\right) \twomatrix{P^2-M^2-U_{\Phi\Phi} & -U_{\Phi\phi} \\ -U_{\phi\Phi} & P^2-m^2-U_{\phi\phi}} \column{\Phi^\prime \\ \phi^\prime}  \, .
\end{equation}
With a slight abuse of notation, we will simply denote the extended version of the $U'$ matrix in Eq.~\ref{eq:lagrangianUV} by ${\bf U}$,
\beq
{\bf U} = \twomatrix{U_{\Phi\Phi} & U_{\Phi\phi} \\ U_{\phi\Phi} & U_{\phi\phi}}\,. \label{Umatrix}
\eeq
For scalar fields, this is the same as the $U$ matrix which can be substituted and evaluated in the UOLEA of Eq.~\ref{eq:universallagrangian} in the usual way~\footnote{The quadratic term has to be put in a form such that $\Phi$ and $\phi$ have the same $c_s$. We will see an example of how this is done in Section~\ref{sec:electroweaktriplet}.}. 

The procedure above is equivalent to the well-known background field method for calculating the one-loop 1PI effective action in the UV theory~\footnote{For more details on the background field method, see for example Ref.~\cite{backgroundfieldmethod}.}. It is easily seen that $U_{\Phi\Phi}$, $U_{\Phi\phi,\phi\Phi}$, $U_{\phi\phi}$ have the diagrammatic interpretation as contributions from heavy-heavy, heavy-light, light-light loops, respectively. In the present context of matching an UV theory to a local EFT, it is helpful to consider the corresponding diagrammatic contributions to the 1PI amplitudes in the EFT (which are one-light-particle-irreducible in the UV theory).
\begin{itemize}
\item Heavy-heavy loops in the UV theory do not appear in the EFT. Instead, they are encoded in local effective operators involving $\phi$. Intuitively, the $\Phi$ loop is effectively shrunk to a point at low energy. As a result, the UOLEA terms involving only $U_{\Phi\Phi}$ correspond to the usual heavy-heavy contributions to one-loop matching~\cite{HLM,UOLEA}. This is illustrated in Fig.~\ref{fig:diagramoneloopheavyonly}.
\item Light-light loops are the same in the UV theory and in the EFT. Thus, terms involving only $U_{\phi\phi}$ do not contribute to matching.
\item Heavy-light loops in the UV theory are depicted schematically in Fig.~\ref{fig:diagramoneloop} and correspond to two pieces in the EFT -- one-loop diagrams obtained by shrinking the $\Phi$ propagators in the loop to a point, and tree diagrams obtained by replacing the entire loop by an effective contact interaction. The former include one-loop EFT diagrams with tree-generated operator insertions, which are part of the 1PI effective action but not part of $\cal L_\text{eff}$ (which only contains local operators). The latter correspond to one-loop-generated operators used at tree level, and constitute the heavy-light contributions to matching that we aim to evaluate with the functional approach. However, the UOLEA terms involving the off-diagonal $U_{\Phi\phi,\phi\Phi}$ are associated with universal coefficients $f_N$ which contain both pieces discussed above. We will use the following procedure to identify the first piece which we subtract from $f_N$ to obtain the subtracted versions $\left(f_N\right)_\text{sub}$.
\end{itemize}

\subsection{Subtracted Universal Coefficients}
\label{sec:subtraction}

An intermediate step in deriving the UOLEA is to compute the integrals over propagators and their momentum derivatives to obtain the universal coefficients $f_N$, where $N = 1, \text{...}, 19$~\cite{UOLEA}. For example, for $f_7$ we have
\begin{equation*}
f_7^{ij} = -\frac{1}{2d} \int dq\,d\xi\, \Delta^2_{\xi,i}(\partial^2\Delta_{\xi,j})m_i^2 \, ,
\end{equation*}
where we have introduced shorthand notation $\int dqd\xi \equiv \int \frac{d^d q}{(2\pi)^d} \int d\xi$, $\Delta_{\xi,i} \equiv 1/(q^2 - \xi m^2_i)$, $\partial_\mu \equiv \frac{\partial}{\partial q^\mu}$, $m_i \equiv \mathcal{M}_{ii}$, and we set $\xi = 1$ after integrating. 

To obtain the {\it subtraction term} $\Delta f_N$ corresponding to $f_N$, we proceed as follows. First, perform the $\xi$ integral. For this to be done easily in closed form, integration by parts on $q$ may be necessary. Then, we replace partial derivatives of the form 
\begin{equation}
\partial_{\mu_1}\dots\partial_{\mu_n} \frac{1}{q^2 - m^2} \quad \longrightarrow \quad \frac{\partial}{\partial k^{\mu_1}}\dots\frac{\partial}{\partial k^{\mu_n}} \left.\frac{1}{(q+k)^2 - m^2}\right|_{k\to 0} \, ,
\end{equation}
and move $\frac{\partial}{\partial k}$ outside of the $q$-integral | this step is reminiscent of the extraction of external momentum dependence of amplitudes in diagrammatic matching and allows us to isolate the leading structures corresponding to heavy propagators shrunk to a point. The latter is achieved by expanding the {\it heavy} propagators in the integrand as
\begin{align*}
\frac{1}{q^2-m^2} &= -\frac{1}{m^2} - \frac{q^2}{m^4} - \text{...} \, , \\
\frac{1}{(q+k)^2 - m^2} &= -\frac{1}{m^2} - \frac{(q+k)^2}{m^4} - \text{...} \, ,
\end{align*}
etc.\ while keeping the light propagators intact. This is the key step that allows us to extract contributions to the 1PI effective action from one-loop EFT diagrams with tree-generated local operator insertions, namely the first piece of heavy-light loops discussed in the last bullet above (first diagram on the right in the example of Fig.~\ref{fig:diagramoneloop}).

Finally, we get the subtraction term $\Delta f_N$ by evaluating the $q$-integral, taking $k$-derivatives as required, for terms in the $1/m^2$ expansion up to the desired order. The {\it subtracted coefficient} is then
 \begin{equation}
 (f_N)_\text{sub} = f_N - \Delta f_N . 
 \end{equation}
 The interpretation of this equation should be clear from our discussion in the last bullet above. 
 $f_N$ is essentially the full expression of the heavy-light loops in the UV theory (left side of Fig.~\ref{fig:diagramoneloop}), which is matched onto the sum of two pieces in the EFT (right side of Fig.~\ref{fig:diagramoneloop}) |  $\Delta f_N$, corresponding to tree-generated operators used in one-loop diagrams, and $(f_N)_\text{sub}$, corresponding to loop-generated operators used at tree-level. Some sample calculations of the subtraction terms $\Delta f_N$ can be found in Appendix~\ref{app:usefulformulas}.

\section{An electroweak triplet scalar example}
\label{sec:electroweaktriplet}

To illustrate the above method at work, we consider a simple extension of the SM by a heavy electroweak scalar triplet, which generates dimension-6 operators involving the light Higgs doublet. The scalar sector of the model is given by
\begin{align}
\mathcal{L} \,\supset\, & |D_\mu H|^2 - m^2|H|^2 - \lambda|H|^4 + \frac{1}{2} (D_\mu\Phi^a)^2 - \frac{1}{2}M^2\Phi^a\Phi^a -\frac{1}{4}\lambda_\Phi(\Phi^a\Phi^a)^2 \nonumber\\
&+ \kappa H^\dagger\sigma^a H \Phi^a - \eta |H|^2 \Phi^a\Phi^a \, ,
\label{eq:Ltriplet}
\end{align}
where $H$ is the light Higgs doublet with hypercharge $Y_H = 1/2$ and mass squared $m^2<0$, $\Phi$ is the heavy $SU(2)_L$ triplet with null hypercharge, and the covariant derivatives are defined accordingly. 

The heavy-heavy loop contributions to the one-loop effective Lagrangian have already been worked out in~\cite{HLM} using the functional approach. We shall focus on the mixed heavy-light loop contributions previously obtained by Feynman diagram methods in Refs.~\cite{skiba,skibatasi, AKS}. In particular, we will work out explicitly the scalar sector contributions to a subset of effective operators generated in this model, and discuss extensions needed to fully incorporate the gauge sector contributions.

\subsection{The scalar sector}

To begin with, we separate both the heavy and light scalar fields into classical backgrounds and quantum fluctuations,
\begin{equation}
 \vec{\Phi} = \vec{\Phi}_c + \vec{\Phi}^\prime \quad , \quad H = H_c + H^\prime\, .
\end{equation}
It will be convenient to define $\tilde{H} \equiv i\sigma^2 H^*$, which transforms the same way as $H$ under $SU(2)_L$. Collecting the quadratic terms, we obtain 
\begin{equation}
\mathcal{L}_\text{quad.} = \frac{1}{2}(\vec{\Phi}^{\prime T}, H^{\prime\dagger}, \tilde{H}^{\prime\dagger}) 
\twomatrix{ P^2 - M^2 - U_{\Phi\Phi} & -\left({\bf U}_{\Phi H}\right)_{1\times2}  \\ -\left({\bf U}_{H\Phi}\right)_{2\times1} & \left(P^2 - m^2 -{\bf U}_{HH}\right)_{2\times2} } 
\column{\vec{\Phi}^{\prime} \\ H^{\prime} \\ \tilde{H}^{\prime}} \, ,
\label{eq:quad}
\end{equation}
where we have labeled the sizes of ${\bf U}_{\Phi H}$, ${\bf U}_{H \Phi}$ and ${\bf U}_{HH}$ matrices in the space of the $\left(\vec{\Phi}^\prime, H^\prime, \tilde{H}^\prime\right)$ multiplet, and made the gauge indices implicit. Note that the separation of the complex doublet $H$ into $H$ and $\tilde{H}$ is essential for all the fields in the multiplet to have a common $c_s=\half$.

The explicit form of the extended ${\bf U}$ matrix for the scalar sector,
\beq
{\bf U} = \twomatrix{U_{\Phi\Phi} & \left({\bf U}_{\Phi H}\right)_{1\times2} \\
\left({\bf U}_{H\Phi}\right)_{2\times1} & \left({\bf U}_{HH}\right)_{2\times2} }
\label{eq:Uscalar}
\eeq
can be derived from Eq.~(\ref{eq:Ltriplet}), which reads
\begin{align}
U_{\Phi\Phi} &= 2\eta |H_c|^2 \identity_3 + \lambda_\Phi\left[ (\vec{\Phi}^T_c\vec{\Phi}_c)\identity_3 + 2\vec{\Phi}_c\vec{\Phi}_c^T \right] \, , \\
{\bf U}_{\Phi H} &= \left( -\kappa H_c^\dagger \vec{\sigma} + 2\eta \vec{\Phi}_c H_c^\dagger \,,\, \kappa \tilde{H}_c^\dagger \vec{\sigma} + 2\eta \vec{\Phi}_c \tilde{H}_c^\dagger \right) \,, \\
{\bf U}_{H\Phi} &= \column{ -\kappa \vec{\sigma}^T H_c + 2\eta H_c \vec{\Phi}_c^T \\ \kappa \vec{\sigma}^T \tilde{H}_c + 2\eta \tilde{H}_c \vec{\Phi}_c^T } = \left({\bf U}_{\Phi H}\right)^{\dagger T} \,,\\
{\bf U}_{HH} &= \twomatrix{ U_{HH} & U_{H\tilde{H}} \\ U_{\tilde{H}H} & U_{\tilde{H}\tilde{H}} } \, ,
\end{align}
where
\begin{align}
U_{H\tilde{H}} &= 2\lambda H_c\tilde{H}_c^\dagger \, , \\
U_{\tilde{H}H} &= 2\lambda \tilde{H}_c H_c^\dagger \, , \\
U_{HH} &= 2\lambda (|H_c|^2\identity_2 + H_cH_c^\dagger) - \kappa\vec{\Phi}_c^T\vec{\sigma} + \eta(\vec\Phi_c^T\vec\Phi_c)\identity_2 \, , \\
U_{\tilde{H}\tilde{H}} &= 2\lambda(|H_c|^2\identity_2 + \tilde{H}_c\tilde{H}_c^\dagger) + \kappa \vec\Phi_c^T\vec\sigma + \eta(\vec\Phi_c^T\vec\Phi_c)\identity_2 \, .
\end{align}
Note that in our notation, the transpose superscript ``$T$'' is solely meant to turn an $SU(2)_L$ triplet represented by a column vector into the same triplet represented by a row vector | it does not, e.g., transpose ${\bf U}_{\Phi H}$ in the $\left(\vec{\Phi}^\prime, H^\prime, \tilde{H}^\prime\right)$ multiplet space; nor does it take $H$ to $(H^\dagger)^*$. On the other hand, dagger denotes hermitian conjugate, so that $\left({\bf U}_{\Phi H}\right)^\dagger$ becomes a $2\times1$ matrix in the multiplet space. Also, we have used $\identity_3$ and $\identity_2$ to denote identity matrices in $SU(2)_L$ representation space (adjoint and fundamental, respectively).

The background heavy field $\vec\Phi_c$ in the above equations should be substituted by the solution to the classical equation of motion, expanded in terms of local operators in powers of $\frac{1}{M}$,
\begin{equation}
\vec{\Phi}_c = \frac{\kappa}{M^2}H_c^\dagger \vec{\sigma}H_c - \frac{\kappa}{M^4}\Bigl[2\eta|H_c|^2(H_c^\dagger \vec{\sigma} H_c) +D^2(H_c^\dagger \vec{\sigma} H_c)\Bigr] +{\cal O}\left(\frac{1}{M^5}\right) \, ,
\label{eq:subphic}
\end{equation}
where we count $\kappa$ as ${\cal O}(M)$. The two terms displayed above have operator dimensions 2 and 4, respectively, and are sufficient for computing the effective Lagrangian up to dimension 6. We remark that the substitution Eq.~\ref{eq:subphic} is not part of the calculation of the 1PI effective action for the full theory, which involves $\vec\Phi_c$ as well as $H_c$. By setting $\vec\Phi_c$ to the local operator expansion in Eq.~\ref{eq:subphic}, we obtain contributions to the effective Lagrangian for the light fields which correspond to 1-particle-{\it reducible} but 1-light-particle-{\it irreducible} one-loop diagrams.

We may now plug the extended ${\bf U}$ matrix Eq.~\ref{eq:Uscalar} into the UOLEA of Eq.~\ref{eq:universallagrangian}. As alluded to above, we are interested in terms involving ${\bf U}_{\Phi H}$ and ${\bf U}_{H \Phi}$, corresponding to mixed heavy-light contributions to EFT matching at one loop in the scalar sector. Some of these terms also involve $U_{\Phi\Phi}$ and/or ${\bf U}_{HH}$, corresponding to possible additional insertions of background currents attached to heavy and/or light propagators in the heavy-light loop in the diagrammatic language. For illustration purpose, we shall focus on extracting the following dimension-6 operators involving Higgs fields~\footnote{While from the EFT point of view, focusing on a subset of effective operators without specifying the complete operator basis being used leads to ambiguity, in the matching procedure this ambiguity of basis choice can be avoided as long as one keeps track of field and parameter redefinitions.}, 
\begin{equation}
\mathcal{O}_T = \frac{1}{2} \left( H^\dagger \Dfbd H \right)^2 \quad , \quad
\mathcal{O}_H = \frac{1}{2} \left( \partial_\mu|H|^2 \right)^2 \quad , \quad
\mathcal{O}_R = |H|^2|D_\mu H|^2 \, ,
\label{eq:ops}
\end{equation}
where $H^\dagger \Dfbd H = H^\dagger (D_\mu H) -(D_\mu H^\dagger)H$. The subscript $c$ on $H_c$ is dropped for clarity from here on. We will also extract the matching contribution to the Higgs kinetic term $|D_\mu H|^2$, since it necessitates a rescaling of the $H$ field in the EFT, which changes the dimension-6 operator coefficients.

Since no terms with field strengths $G_{\mu\nu}^\prime$ can contribute to the operators of interest, we can drop those from Eq.~\ref{eq:universallagrangian}. Also, terms with more than two $P_\mu=iD_\mu$'s can be dropped. Finally, we note that $f_2^i U_{ii}$ cannot involve the off-diagonal ${\bf U}_{\Phi H,H \Phi}$, while terms with more than three $U$'s but no $P_\mu$'s cannot contribute to the operators of interest\footnote{For these terms, we need two covariant derivatives coming from $\vec\Phi_c$, and it is then easily seen that the operator dimension exceeds 6.}. We are thus left with the following terms in the UOLEA (recall that $c_s=\half$),
\begin{align}
{\cal L}^{\text{eff}}_{\text{1-loop}}[H] \supset - \frac{i}{2} & \Bigg[ 
f_{4}^{ij} \text{tr}U_{ij}^2  +f_{7}^{ij} \text{tr}[P_{\mu}, U_{ij}]^2 +f_{8}^{ijk} \text{tr}(U_{ij}U_{jk}U_{ki})   
+ f_{11}^{ijk} \text{tr}(U_{ij} [P_{\mu},U_{jk}][P_{\mu},U_{ki}])  \nonumber \\
& + f_{17}^{ijkl} \text{tr}(U_{ij} U_{jk}  [P_{\mu},U_{kl}][P_{\mu},U_{li}]) +f_{18}^{ijkl}  \text{tr}(U_{ij} [P_{\mu},U_{jk}] U_{kl} [P_{\mu},U_{li}] )
\Bigg] \, ,
\label{eq:UOLEAsubset}
\end{align}
where we have explicitly written out ``tr'' to indicate traces should be taken over gauge indices ($SU(2)_L$ adjoint for $\vec\Phi$, and $SU(2)_L$ fundamental for $H$, $\tilde H$) that have been made implicit in this section. Since there are three non-degenerate masses, the latin indices $i,j,k,l$ range over the values $1,2,3$ that we shall label as $\Phi, H, \tilde{H}$ for clarity.
The calculation may be separated into two parts, starting with the evaluation of the matrix traces followed by the subtraction procedure on the universal coefficients. Some useful formulae for these purposes are collected in Appendix~\ref{app:usefulformulas}. For the first part, we obtain the result
\begin{align}
{\cal L}^{\text{eff}}_{\text{1-loop}}[H] &\supset -i \Bigg\{ \left(f_4^{\Phi H} + f_4^{H \Phi} \right)_\text{sub} \frac{\kappa^2\eta}{M^4}\left(- 8\mathcal{O}_T - 16 \mathcal{O}_R \right)   \,   \nonumber \\
& + \left(f_7^{\Phi H} + f_7^{H\Phi}\right)_\text{sub} \left[ -3\kappa^2 |D_\mu H|^2 + \frac{\kappa^2\eta}{M^2}\left(4\mathcal{O}_H + 4\mathcal{O}_R\right) \right] \, \nonumber \\
&+ \left(f_8^{\Phi H H} + 2f_8^{HH\Phi}\right)_\text{sub} \frac{\kappa^4}{M^4} (2\mathcal{O}_T +4\mathcal{O}_R) \,  \nonumber \\
&+ \left(f_{11}^{\Phi\Phi H}\right)_\text{sub} \left(-6\kappa^2 \eta\right) \mathcal{O}_R + \left(f_{11}^{H\Phi\Phi}\right)_\text{sub}\left(-12\kappa^2\eta\right)\mathcal{O}_H \, \nonumber \\
&+ \left(f_{11}^{HH\Phi}\right)_\text{sub} \kappa^2\left[ \left(\frac{\kappa^2}{M^2}-2\lambda\right)\mathcal{O}_T - \frac{\kappa^2}{M^2}\mathcal{O}_H + \left(\frac{\kappa^2}{M^2} - 10\lambda\right) \mathcal{O}_R \right] \, \nonumber \\
&+ \left(f_{11}^{\Phi HH}\right)_\text{sub} \kappa^2 \left[ \left(- \frac{2\kappa^2}{M^2} + 4\lambda\right)\mathcal{O}_T - 20\lambda\mathcal{O}_H - \frac{4\kappa^2}{M^2}\mathcal{O}_R \right] \, \nonumber \\
&+ \left(f_{17}^{\Phi H \Phi H}\right)_\text{sub} (-6\kappa^4 )\mathcal{O}_R + \left(f_{17}^{H\Phi H\Phi}\right)_\text{sub} \kappa^4 \left( -\mathcal{O}_H - 4\mathcal{O}_R\right) \, \nonumber \\
&+ \left(f_{18}^{\Phi H \Phi H}\right)_\text{sub} \kappa^4 \left( - 10\mathcal{O}_H + 8 \mathcal{O}_R \right) 
\Bigg\} \, . 
\label{eq:traceresult}
\end{align}
The terms are conveniently grouped according to their contributions from each UOLEA term, indicating how the dimension-6 operators originate from these universal building blocks. Note that at this stage, no distinction between $H$ and $\tilde H$ is needed in the indices of the universal coefficients $f_N$ since they have the same mass $m$. On the other hand, all appearances of $\tilde H$ in the operators can be removed in favor of $H$, as shown in Appendix~\ref{app:usefulformulas}.

Next we compute the subtracted universal coefficients $(f_N)_\text{sub}=f_N-\Delta f_N$ according to the procedure described in Subsection~\ref{sec:subtraction}. We shall work up to $\frac{1}{M}$ orders that are sufficient to obtain the operator coefficients of interest, using the $\overline{\text{MS}}$ scheme throughout with renormalization scale $\mu$. The unsubtracted universal coefficients $f_N$ are readily available from~\cite{UOLEA} (except for $f_4$, which can nevertheless be easily calculated),
\begin{small}
\begin{align}
&f_4^{\Phi H} + f_4^{H \Phi} = \frac{i}{16\pi^2} \left(1-\log\frac{M^2}{\mu^2}\right) + \mathcal{O}\left(\frac{1}{M^2}\right)\, , \nonumber \\
&f_7^{\Phi H} + f_7^{H\Phi} = \frac{i}{16\pi^2} \left(-\frac{1}{2M^2}\right) +\mathcal{O}\left(\frac{1}{M^4}\right)\, , \nonumber \\
&f_8^{\Phi HH} + 2f_8^{HH\Phi} = \frac{i}{16\pi^2} \frac{1}{M^2}\left(1-\log\frac{M^2}{m^2}\right) +\mathcal{O}\left(\frac{1}{M^4}\right)\, , \nonumber \\
&f_{11}^{\Phi\Phi H} = \frac{i}{16\pi^2} \frac{1}{2M^4} +\mathcal{O}\left(\frac{1}{M^6}\right)\, , \qquad
f_{11}^{H\Phi\Phi} = \frac{i}{16\pi^2} \frac{1}{3M^4} +\mathcal{O}\left(\frac{1}{M^6}\right)\, , \nonumber \\
&f_{11}^{HH\Phi} = \frac{i}{16\pi^2} \frac{1}{M^4}\left(-\frac{5}{2} +\log\frac{M^2}{m^2}\right) +\mathcal{O}\left(\frac{1}{M^6}\right)\, , \nonumber \\
&f_{11}^{\Phi HH} = \frac{i}{16\pi^2} \left(\frac{1}{6m^2M^2} -\frac{1}{3M^4}\right) +\mathcal{O}\left(\frac{1}{M^6}\right)\, , \nonumber \\
&f_{17}^{\Phi H \Phi H} = \frac{i}{16\pi^2}  \left(-\frac{1}{6m^2M^4} +\frac{2}{3M^6}\right) +\mathcal{O}\left(\frac{1}{M^8}\right)\,, \nonumber \\
&f_{17}^{H \Phi H \Phi} = \frac{i}{16\pi^2}\frac{1}{M^6} \left(\frac{17}{6}-\log\frac{M^2}{m^2}\right) +\mathcal{O}\left(\frac{1}{M^8}\right)\, , \nonumber \\
&f_{18}^{\Phi H \Phi H} = \frac{i}{16\pi^2} \left(-\frac{1}{12m^2M^4} +\frac{1}{4M^6}\right) +\mathcal{O}\left(\frac{1}{M^8}\right)\,.
\label{eq:intermediatesub1}
\end{align}
\end{small}
The subtraction terms $\Delta f_N$ are obtained as follows,
\begin{small}
\begin{align}
&\Delta\left(f_4^{\Phi H} + f_4^{H \Phi}\right) = \mathcal{O}\left(\frac{1}{M^2}\right)\, , \nonumber \\
&\Delta\left(f_7^{\Phi H} + f_7^{H\Phi}\right) = \mathcal{O}\left(\frac{1}{M^4}\right)\, , \nonumber \\
&\Delta\left(f_8^{\Phi HH} + 2f_8^{HH\Phi}\right) = \frac{i}{16\pi^2} \frac{1}{M^2}\log\frac{m^2}{\mu^2} +\mathcal{O}\left(\frac{1}{M^4}\right)\, , \nonumber \\
&\Delta\left(f_{11}^{\Phi\Phi H}\right) = \mathcal{O}\left(\frac{1}{M^6}\right)\, , \qquad
\Delta\left(f_{11}^{H\Phi\Phi}\right) = \mathcal{O}\left(\frac{1}{M^6}\right)\, , \nonumber \\
&\Delta\left(f_{11}^{HH\Phi}\right) = \frac{i}{16\pi^2} \frac{1}{M^4} \left(-\log\frac{m^2}{\mu^2}\right) +\mathcal{O}\left(\frac{1}{M^6}\right)\, , \nonumber \\
&\Delta\left(f_{11}^{\Phi HH}\right) = \frac{i}{16\pi^2} \left(\frac{1}{6m^2M^2} +\frac{1}{6M^4}\right) +\mathcal{O}\left(\frac{1}{M^6}\right)\, , \nonumber \\
&\Delta\left(f_{17}^{\Phi H \Phi H}\right) = \frac{i}{16\pi^2} \left(-\frac{1}{6m^2M^4} -\frac{1}{3M^6}\right) +\mathcal{O}\left(\frac{1}{M^8}\right)\, , \nonumber \\
&\Delta\left(f_{17}^{H\Phi H\Phi}\right) = \frac{i}{16\pi^2}\frac{1}{M^6} \log\frac{m^2}{\mu^2} +\mathcal{O}\left(\frac{1}{M^8}\right)\, , \nonumber \\
& \Delta\left(f_{18}^{\Phi H \Phi H}\right) = \frac{i}{16\pi^2} \left(-\frac{1}{12m^2M^4} -\frac{1}{6M^6}\right) +\mathcal{O}\left(\frac{1}{M^8}\right)\,.
\label{eq:intermediatesub2}
\end{align}
\end{small}
We therefore arrive at the subtracted universal coefficients, which read
\begin{small}
\begin{align}
&\left(f_4^{\Phi H} + f_4^{H \Phi}\right)_\text{sub} = \frac{i}{16\pi^2} \left(1-\log\frac{M^2}{\mu^2}\right) + \mathcal{O}\left(\frac{1}{M^2}\right) \, , \nonumber \\
&\left(f_7^{\Phi H} + f_7^{H\Phi}\right)_\text{sub} = \frac{i}{16\pi^2} \left(-\frac{1}{2M^2}\right) +\mathcal{O}\left(\frac{1}{M^4}\right)\, , \nonumber \\ 
&\left(f_8^{\Phi HH} + 2f_8^{HH\Phi}\right)_\text{sub} = \frac{i}{16\pi^2} \frac{1}{M^2}\left(1-\log\frac{M^2}{\mu^2}\right) +\mathcal{O}\left(\frac{1}{M^4}\right) \, , \nonumber \\
&\left(f_{11}^{\Phi\Phi H}\right)_\text{sub} = \frac{i}{16\pi^2} \frac{1}{2M^4} +\mathcal{O}\left(\frac{1}{M^6}\right)\, , \qquad 
\left(f_{11}^{H\Phi\Phi}\right)_\text{sub} = \frac{i}{16\pi^2} \frac{1}{3M^4} +\mathcal{O}\left(\frac{1}{M^6}\right) \, , \nonumber \\
&\left(f_{11}^{HH\Phi}\right)_\text{sub} = \frac{i}{16\pi^2} \frac{1}{M^4}\left(-\frac{5}{2} +\log\frac{M^2}{\mu^2}\right) +\mathcal{O}\left(\frac{1}{M^6}\right)\, , \nonumber \\
&\left(f_{11}^{\Phi HH}\right)_\text{sub} = \frac{i}{16\pi^2} \left(-\frac{1}{2M^4}\right) +\mathcal{O}\left(\frac{1}{M^6}\right) \, , \nonumber \\
&\left(f_{17}^{\Phi H \Phi H}\right)_\text{sub} = \frac{i}{16\pi^2}\frac{1}{M^6} +\mathcal{O}\left(\frac{1}{M^8}\right)\,, \nonumber \\
&\left(f_{17}^{H\Phi H\Phi}\right)_\text{sub} = \frac{i}{16\pi^2}\frac{1}{M^6} \left(\frac{17}{6}-\log\frac{M^2}{\mu^2}\right) +\mathcal{O}\left(\frac{1}{M^8}\right) \, , \nonumber \\
&\left(f_{18}^{\Phi H \Phi H}\right)_\text{sub} = \frac{i}{16\pi^2}\frac{5}{12M^6} + \mathcal{O}\left(\frac{1}{M^8}\right) \, .
\label{eq:subtractionresult}
\end{align}
\end{small}
Note that while the unsubtracted coefficients and subtraction terms can individually depend on the light SM Higgs mass squared $m^2$, all the $m^2$ dependences drop out in the subtracted universal coefficients, as expected. In particular, the $m^2$ appearing in logarithms in Eqs.~\ref{eq:intermediatesub1} and \ref{eq:intermediatesub2} do not concern us even though $m^2<0$. In general, local effective operator coefficients obtained from matching cannot depend on IR physics~\cite{decoupling}, including electroweak symmetry breaking induced by a negative $m^2$ as in the SM.

Putting together the trace result of Eq.~\ref{eq:traceresult} and the subtracted universal coefficients (\ref{eq:subtractionresult}), we obtain the final expression for the mixed heavy-light one-loop contributions to the dimension-6 operators (\ref{eq:ops}) and the $H$ kinetic term: 
\begin{align}
{\cal L}^{\text{eff}}_{\text{1-loop}}[H] &\supset \frac{1}{16\pi^2}\frac{3\kappa^2}{2M^2}|D_\mu H|^2 + \frac{1}{16\pi^2}\frac{\kappa^2}{M^4}\left[\left(\frac{\kappa^2}{2M^2} -8\eta + 3\lambda \right)\mathcal{O}_T  \, \right. \nonumber \\
& \left. \quad\quad\quad + \left(- \frac{9\kappa^2}{2M^2} -6\eta + 10\lambda \right)\mathcal{O}_H+ \left(- \frac{21\kappa^2}{2M^2} -21\eta + 25\lambda \right)\mathcal{O}_R \right] \, ,
\label{eq:tripletresult}
\end{align}
where we have set the matching scale $\mu=M$~\footnote{In other words, in Eq.~\ref{eq:tripletresult} we report effective operator coefficients renormalized at $\mu=M$ in the $\overline{\text{MS}}$ scheme. It is straightforward to recover the $\mu$ dependence from Eqs.~\ref{eq:traceresult} and~\ref{eq:subtractionresult}, which contains information of the running of operator coefficients.}. Due to the presence of the first term in the above equation, when working with the EFT, one may wish to redefine the $H$ field so that its kinetic term is canonically normalized,
\begin{equation}
H \to \left(1-\frac{1}{16\pi^2}\frac{3\kappa^2}{4M^2}\right) H \quad \Rightarrow \quad \mathcal{L}_\text{eff} \supset \left(1+\frac{1}{16\pi^2}\frac{3\kappa^2}{2M^2}\right) |D_\mu H|^2 \to |D_\mu H|^2,
\end{equation}
up to one loop order. After this rescaling, the one-loop level dimension-6 operator coefficients receive extra contributions from the tree-generated operators~\cite{AKS},
\beq
{\cal L}^{\text{eff}}_{\text{tree}}[H] \supset \frac{\kappa^2}{M^4} (\mathcal{O}_T +2\mathcal{O}_R) \to \left(1-\frac{1}{16\pi^2}\frac{3\kappa^2}{M^2}\right)\frac{\kappa^2}{M^4} (\mathcal{O}_T +2\mathcal{O}_R)
\eeq
As a result, we obtain
\begin{align}
{\cal L}^{\text{eff}}_{\text{1-loop}}[H] &\supset  \frac{1}{16\pi^2}\frac{\kappa^2}{M^4}\left[\left( - \frac{5\kappa^2}{2M^2} -8\eta + 3\lambda\right)\mathcal{O}_T  \, \right. \nonumber \\
& \left. \quad\quad\quad + \left( - \frac{9\kappa^2}{2M^2} -6\eta + 10\lambda\right)\mathcal{O}_H+ \left(- \frac{33\kappa^2}{2M^2} -21\eta + 25\lambda \right)\mathcal{O}_R \right] \, .
\end{align}
These expressions obtained using the UOLEA agree with previous results in the literature~\cite{skiba,skibatasi,AKS}.

\subsection{Extending to the gauge sector} 
\label{sec:vector}

We now turn to the mixed heavy-light contributions including quantum fluctuations of the electroweak gauge vector bosons. This decomposition involves gauge-fixing the quantum part while maintaining the gauge invariance of the classical field, as per the background field method~\cite{backgroundfieldmethod}. In the Feynman gauge, we find the following contribution to the quadratic term which has the familiar form,
%
\begin{align}
\mathcal{L}_\text{quad.} \supset &\, \frac{1}{2}(\vec{\Phi}^{\prime T}, H^{\prime\dagger}, \tilde{H}^{\prime\dagger}, \vec W_\mu^{\prime T}, B'_\mu) \nonumber\\
&\threematrix{ P^2 - M^2 - U_{\Phi\Phi} & -{\bf U}_{\Phi H} & -{\bf U}^{\nu}_{\Phi V}  \\
 -{\bf U}_{H\Phi} & P^2 - m^2 -{\bf U}_{HH} & -{\bf U}^{\nu}_{HV} \\
  -{\bf U}^\mu_{V\Phi} & -{\bf U}^{\mu}_{VH} & -g^{\mu\nu}(P^2-m_V^2)-{\bf U}_{VV}^{\mu\nu}} 
\column{\vec{\Phi}^{\prime} \\ H^{\prime} \\ \tilde{H}^{\prime} \\ \vec W'_\nu \\ B'_\nu} \, ,
\end{align}
where $P_\mu$ is now the covariant derivative with respect to the background gauge fields, and $m_V$ is an IR regulator, to be set to zero at the end of the calculation. The above equation contains to a $5\times5$~${\bf U}$ matrix in the $\left(\vec{\Phi}^\prime, H^\prime, \tilde{H}^\prime, \vec{W}^\prime_\mu, B^\prime_\mu\right)$ multiplet space,
\beq
{\bf U} = \threematrix{U_{\Phi\Phi} & \left({\bf U}_{\Phi H}\right)_{1\times2} & \left({\bf U}^{\nu}_{\Phi V}\right)_{1\times2} \\
\left({\bf U}_{H\Phi}\right)_{2\times1} & \left({\bf U}_{HH}\right)_{2\times2} & \left({\bf U}^{\nu}_{HV}\right)_{2\times2} \\
\left({\bf U}^\mu_{V\Phi}\right)_{2\times1} & \left({\bf U}^{\mu}_{VH}\right)_{2\times2} & \left({\bf U}_{VV}^{\mu\nu}\right)_{2\times2} } \,,
\eeq
with appropriate gauge and Lorentz indices (gauge indices are implicit). The upper-left $3\times3$ block coincides with the ${\bf U}$ matrix involving $\Phi$ and $H$ only from Eq.~\ref{eq:quad} in the previous subsection. The additional elements of the ${\bf U}$ matrix are
\begin{align*}
& {\bf U}^\nu_{\Phi V} = \left(U^\nu_{\Phi W} \,,\, U^\nu_{\Phi B} \right) = \left( ig(D^\nu {\bf\Phi}_c) \,,\, 0 \right) \,, \\
& {\bf U}^\mu_{V\Phi} = \column{U^\mu_{W\Phi} \\ U^\mu_{B\Phi}} = \column{-ig(D^\mu {\bf\Phi}_c) \\ 0} = \left({\bf U}^{\mu}_{\Phi V}\right)^\dagger \,,\\
& {\bf U}^\nu_{HV} = \twomatrix{U^\nu_{HW} & U^\nu_{HB} \\ U^\nu_{\tilde{H}W} & U^\nu_{\tilde{H}B} } = \twomatrix{ -\frac{ig}{2}\vec{\sigma}^T\left(D^\nu H_c\right) & -\frac{ig^\prime}{2}\left(D^\nu H_c\right) \\ -\frac{ig}{2}\vec{\sigma}^T\left(D^\nu \tilde{H}_c\right) & \frac{ig^\prime}{2}\left(D^\nu \tilde{H}_c\right) } \,,\\
& {\bf U}^\mu_{VH} = \twomatrix{U^\mu_{WH} & U^\mu_{W\tilde{H}} \\ U^\mu_{BH} & U^\mu_{B\tilde{H}} } = \twomatrix{ \frac{ig}{2}\left(D^\mu H_c^\dagger\right)\vec{\sigma} & \frac{ig}{2}\left(D^\mu \tilde{H}_c^\dagger\right) \vec{\sigma}\\ \frac{ig^\prime}{2}\left(D^\mu H_c^\dagger\right) & -\frac{ig^\prime}{2}\left(D^\mu \tilde{H}_c^\dagger\right) } = \left({\bf U}^\mu_{HV}\right)^{\dagger T} \,,\\
& {\bf U}^{\mu\nu}_{VV} = \twomatrix{U^{\mu\nu}_{WW} & U^{\mu\nu}_{WB} \\ U^{\mu\nu}_{BW} & U^{\mu\nu}_{BB} } = g^{\mu\nu} 
  \twomatrix{g^2\left[\vec{\Phi_c}\vec{\Phi_c}^T -\left(\vec{\Phi}^T_c\vec{\Phi}_c + \frac{1}{2}|H_c|^2\right)\identity_3\right] & -\frac{gg^\prime}{2}H_c^\dagger\vec{\sigma}H_c \\
  -\frac{gg^\prime}{2}H_c^\dagger\vec{\sigma}^TH_c & -\frac{{g^\prime}^2}{2}|H_c|^2} \,,
\end{align*}
where we have defined the $3\times3$ matrix ${\bf\Phi}$ in the $SU(2)_L$ representation space, to be sandwiched between a row vector and a column vector representing $SU(2)_L$ triplets, with elements $\left({\bf\Phi}_c\right)_{ab} \equiv \Phi_c^e (t_G^e)_{ab}$ where $t_G$ are the $SU(2)_L$ generators in the adjoint representation. 

This extension of the multiplet compared to Eq.~\ref{eq:quad} includes the vector boson quantum fluctuations in the mixed heavy-light one-loop matching computation, and can be used in the UOLEA as in the previous subsection~\footnote{Note that there is an additional piece from the pure Yang-Mills Lagrangian in the ${\bf U}^{\mu\nu}_{VV}$ to be plugged into Eq.~\ref{eq:universallagrangian} that we have omitted in the above equations; see~\cite{HLM}.}. However, in this case there are additional quadratic contributions that contain ``open'' covariant derivatives in the sense that they act on everything to the right (as opposed to appearing in commutators). These extra terms are given by 
%
\beq
\mathcal{L}_\text{quad.} \supset -\frac{1}{2}\left(\vec{\Phi}^{\prime T} , H^{\prime \dagger} , \tilde{H}^{\prime \dagger}  , \vec{W}^{\prime T}_\mu , B^{\prime}_\mu \right) 
 \,  \, {\bf Z} \,  \,
\column{\vec{\Phi}^{\prime} \\ H^{\prime} \\ \tilde{H}^{\prime} \\ \vec{W}^{\prime}_\nu \\ B^{\prime}_\nu} \nonumber \, , 
\eeq
where we have defined the ${\bf Z}$ matrix,
\beq
{\bf Z} = \twomatrix{ \bf{0}_{3\times 3} & \inviscolumn{ P^\nu \left({\bf Z}_{\Phi V}\right)_{1\times2}  \\ \left(P^\nu\right)_{2\times2} \left({\bf Z}_{HV}\right)_{2\times2}  } \\  
\left({\bf Z}_{V\Phi}\right)_{2\times1} P^\mu \quad \left({\bf Z}_{VH}\right)_{2\times2} \left(P^\mu\right)_{2\times2} & {\bf 0_{2 \times 2}} }  \, ,
\eeq
%
with
\begin{align*}
{\bf Z}_{\Phi V} &= \left( Z_{\Phi W} \,,\,  Z_{\Phi B} \right) = \left(g {\bf \Phi_c} \,,\, 0\right) \,,\\
{\bf Z}_{V\Phi} &= \column{Z_{W\Phi} \\ Z_{B\Phi}} = \column{g {\bf \Phi_c} \\ 0} = \left({\bf Z}_{\Phi V}\right)^\dagger \,,\\
{\bf Z}_{HV} &= \twomatrix{Z_{HW} & Z_{HB} \\ Z_{\tilde{H}W} & Z_{\tilde{H}W}} = \twomatrix{-\frac{g}{2}\vec{\sigma}^TH_c & -\frac{g^\prime}{2}H_c \\ -\frac{g}{2}\vec{\sigma}^T\tilde{H}_c & \frac{g^\prime}{2}\tilde{H}_c } \,,\\
{\bf Z}_{VH} &= \twomatrix{Z_{WH} & Z_{W\tilde{H}} \\ Z_{BH} & Z_{B\tilde{H}} } = \twomatrix{-\frac{g}{2}H_c^\dagger\vec{\sigma} & -\frac{g}{2}\tilde{H}_c^\dagger\vec{\sigma} \\ -\frac{g^\prime}{2}H_c^\dagger & \frac{g^\prime}{2}\tilde{H}_c^\dagger } = \left({\bf Z}_{HV}\right)^{\dagger T} \,.
\end{align*}

Since the ${\bf Z}$ matrix includes open covariant derivatives, they are affected by the initial steps in the CDE method, where a sequence of transformations take
\beq
P_\mu \to P_\mu - q_\mu \to -\tilde{G}_{\rho\mu}\partial^\rho -q_\mu 
\eeq
An implicit assumption in the UOLEA is then that the $U$ matrix does not contain open covariant derivatives. Thus, the presence of the ${\bf Z}$ matrix in $\mathcal{L}_\text{quad.}$ leads to additional terms in the UOLEA formula. Nevertheless, this does not affect the generality of the UOLEA approach. In particular, terms already computed in the UOLEA will always contribute to the open derivative-independent part encapsulated in the $U$ matrix. On the other hand, it requires that the UOLEA formula be extended to fully account for all possible universal terms up to dimension-6. Such an extension is also useful for many other applications of the UOLEA, and will be discussed in detail in a future publication~\cite{workinprogress}.

\section{Conclusion}
\label{sec:conclusion}

We have demonstrated a conceptually simple and transparent method for including mixed heavy-light contributions to integrating out heavy particles using the Universal One-Loop Effective Action (UOLEA). This procedure requires separating both heavy and light fields into classical background and quantum fluctuation parts. The $U$ matrix of the quadratic term in the action is consequently extended to include these quantum fluctuations for both heavy and light fields, forming a combined multiplet. The traces over matrices in the UOLEA are then evaluated as usual, with the heavy-light contributions treated together with the heavy-heavy ones.

An additional step is needed for the mixed heavy-light terms. The universal coefficients must be replaced by their subtracted versions since they encapsulate the 1PI effective action when all fields are separated into classical and quantum pieces; in other words, in the low-energy EFT they contain parts where tree-generated operators are used in one-loop diagrams, which should not be included in local effective operator coefficients. 
We developed a subtraction algorithm, which can be used at an intermediate step in the CDE derivation of the UOLEA universal coefficients, resulting in the subtracted universal coefficients.

As an example, we considered integrating out a heavy electroweak triplet scalar coupling to a light Higgs doublet. This model has the benefit of convenient comparison with available results and recent discussion in the literature. We derived the $U$ matrix for the scalar sector of the theory and the relevant subtracted universal coefficients to obtain dimension-6 Higgs operators, illustrating the procedure step-by-step. The intermediate results give insight into how the universal structures of the UOLEA form the building blocks of the resulting Wilson coefficients. 

Vector gauge bosons may also contribute to the mixed heavy-light calculation. We derived the extended $U$ matrix to include scalar-vector interactions, and found an additional piece containing covariant derivatives that act openly to the right, that we called the ${\bf Z}$ matrix. This additional ${\bf Z}$ matrix does not form part of the pre-evaluated UOLEA, since the UOLEA in Ref.~\cite{HLM, UOLEA} is the general result of evaluating only the $U$ matrix with the implicit assumption that it does not contain open covariant derivatives. Such new structures involving covariant derivatives may arise also in other applications of the general framework,
and will be added to the UOLEA in the future~\cite{workinprogress}.

To summarise, we have outlined a procedure for including mixed heavy-light contributions in the functional approach to matching. In particular, we have shown how this relates to the UOLEA that encapsulates the general results and structure of such one-loop path integral computations, and we have found additional structures that may also in principle be universal.
\\
\\
\\
 {\bf Note added:} {\it As this work was being finalised for submission, Ref.~\cite{HLM3} appeared where they also develop the CDE to include mixed heavy-light contributions to one-loop matching. Their approach, while similar in spirit, differs from ours in their subtraction procedure and does not make use of the UOLEA for mixed heavy-light matching. }

\section*{Acknowledgements}
We are grateful for discussions with and encouragement from John Ellis. ZZ thanks James~D.~Wells for useful discussions. JQ and TY would also like to acknowledge discussions with Marco Nardecchia and TY is grateful to King's College London for hospitality while this work was being completed. The work of SE and ZZ was supported in part by the U.S. Department of Energy under grant DE-SC0007859. 
The work of JQ was supported by the STFC Grant ST/L000326/1. 
The work of TY was supported by a Research Fellowship from Gonville and Caius College, Cambridge.

\begin{appendices}
\section{Useful Formulae}
\label{app:usefulformulas}

\subsection{Identities}
In this appendix we group together some formulae for the derivations above. As usual, we use the definition that $\Htil \equiv i \sigma_2 H^*$. We can make use of the following identities to eliminate $\tilde{H}$ in the final results for the operators in favor of $H$:
\beq
\Hdtil \Htil = \Hd H, ~~~ \Hdtil \sigma^a \Htil = - \Hd \sigma^a H, ~~~\Hdtil \sigma^a \sigma^b \Htil = \Hd \sigma^b \sigma^a H, ~~~\text{etc.}\ ,
\eeq
\beq
\Hdtil(\Dm \Htil) = (\Dm \Hd) H,~~~ \Hdtil \sigma^a (\Dm \Htil) = - ( \Dm \Hd) \sigma^a H, ~~~ \text{etc.} \ ,
\eeq

We also make extensive use of the following well-known integral identities in the computation of the subtracted universal coefficients:
\begin{align}
&\int dq \ \frac{1}{q^2-m^2} = \frac{i}{16\pi^2}m^2 \left(\Dv + 1 - \log\frac{m^2}{\mu^2} \right) \ ,\\ 
&\int dq \ \frac{1}{(q^2-m^2)^2} = \frac{i}{16\pi^2} \left(\Dv - \log\frac{m^2}{\mu^2} \right) \ ,\\  
&\int dq \ \frac{1}{(q^2-m^2)(q^2-m'^2)}= \frac{i}{16\pi^2} \left( \Dv + 1 - \frac{m^2 \log \frac{m^2}{\mu^2}-m'^2 \log \frac{m'^2}{\mu^2}}{m^2-m'^2} \right) \ ,\\
& \frac{1}{2d} \frac{\partial}{\partial k_\mu} \frac{\partial}{\partial k^\mu}\int dq \ \frac{1}{(q^2-m^2)((q+k)^2-m^2)}= \frac{i}{16\pi^2} \frac{1}{6m^2} \ , 
\end{align}
where $\int dq \equiv \int \frac{d^d q}{(4\pi)^d}$ with $d=4-\epsilon$. The divergent part $\Dv \equiv \frac{2}{\epsilon} - \gamma + \log 4\pi$ is to be dropped in the $\overline{\text{MS}}$ scheme which we adopt.

\subsection{Formulae for computation of subtracted universal coefficients}
Below we present some useful formulae for the computation of the subtracted universal coefficients $f_N$, where $N = 1,\ldots, 19$. In these formulae we define
\beq
\Delta_{\xi, i} \equiv \frac{1}{q^2 - \xi m_i^2}, ~~~ \Delta_i \equiv \frac{1}{q^2 - m_i^2}, ~~~ \partial_\mu \equiv \frac{\partial}{\partial q^\mu},~~~ m_i \equiv \mathcal{M}_{ii} \ ,
\eeq
and we set $\xi=1$ after integrating, as discussed in Section \ref{sec:subtraction}.

In this section we illustrate how the $f_N$ are obtained when there is one heavy field (or one heavy degenerate multiplet) denoted by $h$ with mass $m_h$, and one light field (or one light degenerate multiplet) denoted by $l$ with mass $m_l$. This scenario is readily applied to the scalar triplet example in Section~\ref{sec:electroweaktriplet}, where we can identify $h=\vec\Phi$, $l=(H, \tilde{H})$. We will list the expressions for the $f_N$ combinations that appear in the evaluation of the UOLEA that are useful for the calculation in Section~\ref{sec:electroweaktriplet}. Additional $f_N$ combinations that may appear in other examples, including more general cases where there are nondegenerate heavy and/or light multiplets, can be similarly derived \cite{workinprogress}.

The coefficient $f_4$ can be found starting from the definition given in \cite{UOLEA}:
\begin{align}
f_4^{ij} &=  \int dq \ d \xi \ \Delta_{\xi, i}^2 \  \Delta_{\xi, j} m_i^2 \ ,
\end{align}
such that
\beq
f_4^{hl}+ f_4^{lh} = \int dq \ d \xi \left( \Delta_{\xi, h}^2 \Delta_{\xi, l} m_h^2 + \Delta_{\xi, h}\Delta_{\xi, l}^2 m_l^2\right) = \int dq \ \Delta_h \Delta_l \ .
\eeq

Unlike $f_4$, for $f_7$ and some of the subsequent coefficients discussed here, we can not use the expressions given in \cite{UOLEA}, but rather must start at an intermediate step in the derivation of the UOLEA. 

Here, for $f_7$ we make use of the definition:
\begin{align}
f_7^{ij} &= -\frac{1}{2d} \int dq \ d \xi \ \Delta_{\xi, i}^2 (\partial^2 \Delta_{\xi, j}) m_i^2 \ ,
\end{align}
where we have also made use of $T_{\mu \nu} = g_{\mu \nu} T \ \Rightarrow  \ T = (1/d) g^{\mu \nu} T_{\mu \nu}$, for some tensor $T_{\mu\nu}$. We then find
\begin{align}
\nonumber f_7^{hl} + f_7^{lh} &= - \frac{1}{2d} \int dq \ d \xi \ \left[ \Delta_{\xi, h}^2 (\partial^2 \Delta_{\xi, l}) m_h^2  + \Delta_{\xi, l}^2 (\partial^2 \Delta_{\xi, h}) m_l^2\right]  \ , \\ 
&= - \frac{1}{2d} \int dq \ d \xi \ \left[ \Delta_{\xi, h}^2 (\partial^2 \Delta_{\xi, l}) m_h^2  + \Delta_{\xi, h} (\partial^2 \Delta_{\xi, l}^2) m_l^2\right] = -\frac{1}{2d} \int dq \ \Delta_h(\partial^2 \Delta_l) \ ,
\end{align}
where an integration by parts has been performed on $q$ when going from the first line to the second.

For $f_8$ we make use of the definition:
\begin{align}
f_8^{ijk} &= \int dq \ d \xi \ \Delta_{\xi, i}^2 \Delta_{\xi, j} \Delta_{\xi,k} m_i^2 \ ,
\end{align}
to find
\begin{align} 
 f_8^{hll} + 2 f_8^{llh} &= \int dq \ d \xi \ \left( \Delta_{\xi, h}^2 \Delta_{\xi, l}^2 m_h^2 + 2\Delta_{\xi, l}^3 \Delta_{\xi, h}m_l^2\right) = \int dq \ \Delta_h \Delta_l^2 \ ,
\end{align}
noting these are the only coefficients we need calculate, as $f_8^{ijk} = f_8^{ikj}$.

For $f_{11}$ we use the definition 
\begin{align}
\nonumber f_{11}^{ijk} &= -\frac{1}{2d} \int dq \ d \xi \left[ \Delta_{\xi, i}^2 \Delta_{\xi, j} ( \partial^2 \Delta_{\xi,k}) m_i^2 + \Delta_{\xi, j}^2 \Delta_{\xi, i} ( \partial^2 \Delta_{\xi,k}) m_j^2 + \Delta_{\xi, k}^2 ( \partial^2 \Delta_{\xi, i} \Delta_{\xi,j}) m_k^2\right] \\
\nonumber &= -\frac{1}{2d} \int dq \ d \xi \left[ \Delta_{\xi, i}^2 \Delta_{\xi, j} ( \partial^2 \Delta_{\xi,k}) m_i^2 + \Delta_{\xi, j}^2 \Delta_{\xi, i} ( \partial^2 \Delta_{\xi,k}) m_j^2 +  \Delta_{\xi, i} \Delta_{\xi,j}( \partial^2 \Delta_{\xi, k}^2) m_k^2\right]  \\
&= -\frac{1}{2d} \int dq \ \Delta_i \Delta_j ( \partial^2 \Delta_k) \ , 
\end{align}
where to go from the first to the second line we have integrated by parts. We use this to find the four coefficients $f_{11}^{hhl}, ~f_{11}^{lhh},~f_{11}^{llh}, ~ f_{11}^{hll}$. We make use of the fact that $f_{11}^{ijk} = f_{11}^{jik}$ to reduce the number of coefficients that need be calculated.

Meanwhile for $f_{17}$ we use
\begin{align}
\nonumber f_{17}^{ijkn} = - \frac{1}{2d} \int dq \ d \xi \large[& \Delta_{\xi, i}^2 \Delta_{\xi, j} \Delta_{\xi, k} (\partial^2 \Delta_{\xi, n})m_i^2 + \Delta_{\xi, i} \Delta_{\xi, j}^2 \Delta_{\xi, k} (\partial^2 \Delta_{\xi, n})m_j^2 \\ &+ \Delta_{\xi, i} \Delta_{\xi, j} \Delta_{\xi, k}^2 ( \partial^2 \Delta_{\xi, n}) m_k^2 + \Delta_{\xi, n}^2 \partial^2(\Delta_{\xi, i} \Delta_{\xi, j} \Delta_{\xi, k}) m_n^2 \large] \ ,
\end{align}
which after integration by parts can be rewritten as 
\begin{align}
 f_{17}^{ijkn}= - \frac{1}{2d} \int dq \ \Delta_i \Delta_j \Delta_k ( \partial^2 \Delta_n) \ , 
\end{align}
to find $f_{17}^{hlhl}$ and $f_{17}^{lhlh}$. 

Finally for the coefficient $f_{18}$, we use the definition 
\begin{align}
f_{18}^{ijkn} = - \frac{1}{2d} \int dq \ d \xi \left[ \Delta_{\xi, i}^2 \Delta_{\xi, j} \partial^2(\Delta_{\xi, k} \Delta_{\xi, n}) m_i^2 + \Delta_{\xi, k} \Delta_{\xi, n}^2 \partial^2( \Delta_{\xi, i} \Delta_{\xi, j}) m_n^2\right] \ ,
\end{align}
to find, for example
\begin{align}
\nonumber f_{18}^{hlhl} &= - \frac{1}{2d} \int dq \ d \xi \left[ \Delta_{\xi, h}^2 \Delta_{\xi, l} \partial^2(\Delta_{\xi, h} \Delta_{\xi, l}) m_h^2 + \Delta_{\xi, h} \Delta_{\xi, l}^2 \partial^2( \Delta_{\xi, h} \Delta_{\xi, l}) m_l^2\right] \\
&= - \frac{1}{4d} \int dq \ d \xi \left[ \Delta_{\xi, h}^2 \Delta_{\xi, l} \partial^2(\Delta_{\xi, h} \Delta_{\xi, l}) m_h^2 + \Delta_{\xi, h} \Delta_{\xi, l}^2 \partial^2( \Delta_{\xi, h} \Delta_{\xi, l}) m_l^2\right. \nonumber \\
&\qquad\qquad\qquad\quad \left.+\Delta_{\xi, h} \Delta_{\xi, l} \partial^2(\Delta_{\xi, h}^2 \Delta_{\xi, l}) m_h^2 + \Delta_{\xi, h} \Delta_{\xi, l} \partial^2( \Delta_{\xi, h} \Delta_{\xi, l}^2) m_l^2\right] \nonumber \\
&= -\frac{1}{4d} \int dq \ \Delta_h \Delta_l \partial^2(\Delta_h \Delta_l) \ ,
\end{align}
where to go from the first to the second lines we have integrated by parts. We can make use of $f_{18}^{ijkn}= f_{18}^{nkji}$ to reduce the number of independent coefficients.

\subsubsection{Sample calculation of subtraction terms $\Delta f_{11}^{ll h}$ and $ \Delta f_{11}^{hll}$}

In this subsection we show a sample calculation of the subtraction terms for the coefficients $f_{11}^{ll h}$ and $f_{11}^{hll}$ for the example of the scalar electroweak triplet in Section~\ref{sec:electroweaktriplet}, following the procedure described in Section~\ref{sec:subtraction}. We have chosen these particular coefficients to walk through the calculation as they exhibit interesting dependence on IR parameters.

We first compute the subtraction term $\Delta f_{11}^{ll h}$, given by 
\begin{align}
\nonumber \Delta f_{11}^{llh} &= -\frac{1}{2d}\frac{\partial}{\partial k_\mu}\frac{\partial}{\partial k^\mu} \int dq \ \left[\frac{1}{(q+k)^2-m_h^2}\cdot\frac{1}{(q^2-m_l^2)^2}\right]_\text{expand in $1/m_h^2$} \\ 
\nonumber&= -\frac{1}{2d}\frac{\partial}{\partial k_\mu}\frac{\partial}{\partial k^\mu} \int dq \left[ - \frac{1}{m_h^2} - \frac{(q+k)^2}{m_h^4} + \mathcal{O}\left(\frac{1}{m_h^6}\right) \right]\frac{1}{(q^2-m_l^2)^2} \\
\nonumber&= -\frac{1}{2d}\frac{\partial}{\partial k_\mu}\frac{\partial}{\partial k^\mu} \int dq \left[\frac{-k^2}{m_h^4} \frac{1}{(q^2-m_l^2)^2} + \mathcal{O}\left(\frac{1}{m_h^6}\right) \right] \\
\nonumber&=\frac{1}{m_h^4} \int dq \frac{1}{(q^2-m_l^2)^2} + \mathcal{O}\left(\frac{1}{m_h^6}\right) \\ 
&= \frac{i}{16\pi^2} \frac{1}{m_h^4}\left( \Dv - \log \frac{m_l^2}{\mu^2}\right) + \mathcal{O}\left(\frac{1}{m_h^6}\right)  \ .
\end{align}
We see that our final answer contains an IR-dependent $\log m_l^2$, which cancels off the IR dependence in the $\log$ term in the coefficient $f_{11}^{ll h}$, so that $(f_{11}^{llh})_\text{sub}$ contains no IR sensitivity, as expected.

Now we show the sample computation of the subtraction term $ \Delta f_{11}^{hll}$, which also exhibits IR dependence.
\begin{align}
\nonumber \Delta f_{11}^{hll} &= -\frac{1}{2d}\frac{\partial}{\partial k_\mu}\frac{\partial}{\partial k^\mu} \int dq \ \left[\frac{1}{q^2-m_h^2}\cdot\frac{1}{q^2-m_l^2}\cdot \frac{1}{(q+k)^2-m_l^2}\right]_\text{expand in $1/m_h^2$} \\
\nonumber &=-\frac{1}{2d}\frac{\partial}{\partial k_\mu}\frac{\partial}{\partial k^\mu} \int dq \ \left[ -\frac{1}{m_h^2} - \frac{(q^2-m_l^2) + m_l^2}{m_h^4} + \mathcal{O}\left(\frac{1}{m_h^6}\right)\right]\cdot\frac{1}{q^2-m_l^2}\cdot \frac{1}{(q+k)^2-m_l^2} \\
\nonumber &=-\frac{1}{2d}\frac{\partial}{\partial k_\mu}\frac{\partial}{\partial k^\mu} \int dq \ \Bigg[ -\frac{1}{m_h^2}\cdot\frac{1}{q^2-m_l^2}\cdot \frac{1}{(q+k)^2-m_l^2} \\&\hspace{2cm}- \frac{1}{m_h^4}\left( \frac{1}{(q+k)^2-m_l^2} + \frac{m_l^2}{(q^2-m_l^2)((q+k)^2-m_l^2)}\right)  + \mathcal{O}\left(\frac{1}{m_h^6}\right)\Bigg] \nonumber\\
\nonumber &=-\frac{1}{2d}\frac{\partial}{\partial k_\mu}\frac{\partial}{\partial k^\mu} \int dq \ \left( -\frac{1}{m_h^2} - \frac{m_l^2}{m_h^4}\right)\frac{1}{(q^2-m_l^2)((q+k)^2-m_l^2)} + \mathcal{O}\left(\frac{1}{m_h^6}\right) \\
&= \frac{i}{16\pi^2}\left[ \frac{1}{6 m_l^2 m_h^2} + \frac{1}{6m_h^4} + \mathcal{O}\left(\frac{1}{m_h^6}\right)\right] \,.
\end{align}
Note that $\int dq \frac{1}{(q+k)^2-m_l^2}$ is $k$-independent by our prescription (where shifting the integration variable $q\to q-k$ is allowed, as in the corresponding diagrammatic calculation). We see that the subtraction term for the coefficient contains a term quadratically sensitive to the light mass, which precisely cancels an identical term in the coefficient $f_{11}^{hll}$, so that the final result for $(f_{11}^{hll})_\text{sub}$ contains no IR sensitivity.

\end{appendices}


 \providecommand{\href}[2]{#2}\begingroup\raggedright

\end{document}